\documentclass[doublecol,figures]{per} 
\LogoHeight{2.1cm}

\usepackage{amssymb}
\usepackage[utf8]{inputenc}
\title{Mesoscale Equipartition of kinetic energy in Quantum Turbulence}

\author{J. Salort{$^1$}, P.-E. Roche{$^1$}  \and E. Leveque{$^2$}  }
\institute{
\inst{1} Institut N\'eel, CNRS/UJF,
 BP166, F-38042 Grenoble Cedex 9, France\\
\inst{2}  Laboratoire de Physique, ENS Lyon, CNRS/Universit\'e de Lyon
F-69364 Lyon, France
 }


\abstract{
The turbulence of superfluid helium is investigated numerically at finite temperature. 
Direct numerical simulations are performed with a ``truncated HVBK" model, which combines the continuous description of the Hall-Vinen-Bekeravich-Khalatnikov equations with the additional constraint that this continuous description cannot extend beyond a quantum length scale associated with the mean spacing between individual superfluid vortices. A good agreement is found with experimental measurements of the vortex density. Besides, by varying the turbulence intensity only, it is observed that the inter-vortex spacing varies with the Reynolds number as $Re^{-3/4}$, like the viscous length scale in classical turbulence. In the high temperature limit, Kolmogorov's inertial cascade is recovered, as expected from previous numerical and experimental studies. As the temperature decreases, the inertial cascade remains present at large scales while, at small scales, the system evolves towards a statistical equipartition of kinetic energy among spectral modes, with a characteristic $k^2$ velocity spectrum. The accumulation of superfluid excitations on a range of mesoscales enables the superfluid to keep dissipating kinetic energy through mutual friction with the residual normal fluid, although the later becomes rare at low temperature. It is found that most of the superfluid vorticity can concentrate on these mesoscales at low temperature, while it is concentrated in the inertial range at higher temperature. This observation should have consequences on the interpretation of decaying turbulence experiments, which are often based on vortex line density measurements.
}

\newcommand{\vn}{\mathbf {v}_n}
\newcommand{\vs}{\mathbf {v}_s}
\newcommand{\vns}{\mathbf {v}_{ns}}

\newcommand{\fns}{\mathbf {F}_{ns}}
\newcommand\bom{{\mbox{\boldmath $\omega_s$}}}

\newcommand{\ron}{\rho_n}

\newcommand{\ros}{\rho_s}

\newcommand{\pn}{p_n}
\newcommand{\ps}{p_s}

\begin{document}

\maketitle

\section{Introduction}
 
The turbulence of quantum fluids, such as  Bose Einstein condensates, superfluid helium and neutron stars,
has attracted much attention over the last decade thanks to experimental progresses 
in flow generation and characterisation (e.g. \cite{Henn:2009,Vinen2002}).
In particular, experiments performed with superfluid $^4$He generate turbulence which is intense enough to allow a statistical characterisation of the fluctuations of ``quantum turbulence''.

According to Landau and Tisza, the superfluid state of $^4$He (He~II) can be described as an intimate mixture of two fluids : a viscous \textit{normal} fluid and an inviscid \textit{superfluid} with quantized circulation of velocity. A mutual coupling allows for an exchange of momentum between these two fluid components. The relative fraction of each component strongly depends on the temperature: the normal-fluid vanishes at 0\,K while the superfluid extinguishes at the superfluid transition, around 2.17~K \cite{BarenghiLivre,DonnellyLivreVortices,Vinen2002}.

Like in ordinary fluids, intense turbulence can be generated in He~II by mechanical means, for example by shearing the flow with counter-rotating propellers \cite{Maurer1998}, by imposing an external pressure difference in pipes \cite{Rousset:Pipe1994,Holmes1992,Fuzier2001,RocheVortexSpectrum:EPL2007} or by destabilizing the flow with an obstacle, usually a grid\cite{Niemela:JLTP2005,Salort:PoF2010}. This study addresses this type of turbulence at finite temperature, where the presence of normal-fluid cannot be neglected.

At the largest scales of such turbulent flows, the normal-fluid and the superfluid are nearly locked together as a result of the mutual coupling, which tends to minimize the velocity difference between these two components. Both fluids undergo a common inertial cascade, similar to Kolmogorov cascade in ordinary fluids. This cascade carries kinetic energy from the largest scales of the flow down to smaller scales. Experimental\cite{Maurer1998,Salort:PoF2010} and numerical results\cite{RocheVortexSpectrum:EPL2007}, as well as theoretical arguments\cite{Vinen2002}, support this widely accepted picture.

At smaller scales, it is unclear what happens to the superfluid energy which has cascaded from the largest inertial scales. 
Indeed, the superfluid can loose kinetic energy through mutual coupling with the normal-fluid but this coupling becomes less efficient at low temperature since it is proportional to the normal-fluid fraction. The motivation of this study is to understand the response of the superfluid to dissipate energy and reach a (statistically) stationary state when the normal-fluid density is low.

 High resolution numerical simulations are performed to solve the dynamical equations at four different temperatures, corresponding to significantly different fractions of the normal-fluid and superfluid components. A modified version of the Hall-Vinen-Bekeravich-Khalatnikov (HVBK) model is considered to account for the cut-off length scale associated with the quantization of superfluid vortices.
At low temperature, we observe that a new phenomenon emerges between the inertial length scales and the vortex quantization length scale. In this range of \textit{mesoscales}, the superfluid kinetic energy piles up, which indirectly increases the mutual coupling and compensates for the reduced fraction of normal fluid. The superfluid eventually reaches a stationary state in which the energy distribution
tends towards equipartition with a characteristic $k^2$ spectrum (in the continuation of the $k^{-5/3}$ spectrum related to the Kolmogorov's energy cascade). Interestingly, the amount of superfluid vorticity which accumulates in these \textit{mesoscales} can be much larger than the amount of vorticity held in the inertial scales. This supports a recent prediction\cite{RocheInterpretation:EPL2008} and should be of importance for the interpretation of decaying quantum turbulence experiments, which are all based on measurements of the superfluid vortex density.

\section{A truncated HVBK model}

The  Hall-Vinen-Bekeravich-Khalatnikov model (HVBK)  describes the dynamics of He~II by continuous equations : a Navier--Stokes equation (for the normal-fluid) and an Euler equation (for the superfluid) 
\cite{BarenghiLivre}
The Euler equation is derived by coarse-graining the superfluid field : the details of the vortex tangle are averaged out by smoothing the velocity field on a scale corresponding to the typical inter-vortex spacing $\delta$. Thus, this original model removes all the information about the quantification of vortices and, by construction, does not account physically for the possible propagation of some superfluid excitations (through Kelvin waves along vortices, for instance) at scales smaller than $\delta$. As a first approximation, we neglect these effects and therefore impose that the cut-off scale of the simulation corresponds to the quantum scale $\delta$. 
 A mutual coupling term allows a consistent exchange of momentum between the normal-fluid and the superfluid. The HVBK model has been widely used to simulate quantum fluids (helium and neutron stars) in numerical studies
(e.g. see \cite{Barenghi1988,Barenghi:PRB1992,Henderson:2004,Merahi:2006,Peralta2008,Roche2fluidCascade:EPL2009,Tchoufag:POF2010})For a discussion of models proposed for Bose Einsein condensates, see \cite{Proukakis:2008}. The simulated equations are :
\begin{eqnarray}
\frac{D \vn}{D t} = -\frac{1}{\ron} \nabla \pn  + \frac{\rho_s}{\rho} \fns 
                    + \nu \nabla^2 \vn + {\bf f}_n^{ext},\\
\frac{D \vs}{D t} = -\frac{1}{\ros} \nabla \ps   - \frac{\rho_n}{\rho}\fns
                    + {\bf f}_s^{ext},
\label{eq:main}
\end{eqnarray}

\noindent
where the indices $n$ and $s$ refer to the normal fluid and superfluid 
respectively, ${\bf f}_{n}^{ext}$ and ${\bf f}_{s}^{ext}$ are 
external forcing terms, $\nu$ is the kinematic viscosity ($\nu=\mu /\rho_{n}$), $\rho_{n}$ and $\rho_{s}$ are the normal-fluid
and superfluid densities, $\rho=\rho_n+\rho_s$, $\pn=(\rho_n/\rho)p+\rho_s ST$ and 
$\ps=(\rho_s/\rho) p -\rho_s ST$ are partial pressures, $S$, $T$ and $p$ are
specific entropy, temperature and pressure, and
$\vn$ and $\vs$ satisfy the incompressibility conditions
$\nabla \cdot \vn=0$ and $\nabla \cdot \vs=0$.
The mutual coupling term is :
\begin{equation}
\fns=\frac{B}{2} \frac{\bom}{\vert \bom \vert} \times (\bom \times \vns)
+\frac{B'}{2} \bom \times \vns
\label{eq:fnsHVBK}
\end{equation}

\noindent
where $\vns=\vn-\vs$ is the slip velocity, $\bom=\nabla \times \vs$ is the superfluid vorticity
.
Unless otherwise specified, this mutual coupling was approximated at first order :
\begin{equation}
\fns = -\frac{B}{2} \vert \bom \vert \vns,
\label{eq:fns}
\end{equation}

A simple analytical derivation proposed in \cite{Roche2fluidCascade:EPL2009} (for a slightly different derivation, see \cite{skrbek:JLTP2010} )
 showed that this first-order approximation allows to account for effective viscosity measurements over a large range of temperature in turbulent $^4$He. This supports the use of Eq.\ref{eq:fns} as a reasonable approximation for the mutual coupling.

As already mentioned, the originality of our HVBK modeling consists in preventing the superfluid energy to cascade beyond length scales smaller than an estimated mean-intervortex spacing $\delta$. In other words, the spectral domain is truncated to wave-vectors such that $k\le {2 \pi}/{\delta}$. 
The inter-vortex scale $\delta$ is estimated from the quantum of circulation $\kappa$ around a single superfluid vortex and from the average vorticity :
\begin{eqnarray}
\label{eq:truncation}
\kappa =  \delta ^2 \left(  \frac{1}{V} \cdot \int_V \bom^2 dv   \right) ^{1/2}  = \delta ^2 \left( \overline{\bom ^2} \right) ^{1/2}
\end{eqnarray}


Physically, this truncation procedure is justified as long as the energy cascade on scales smaller than $\delta$ is less efficient than the dissipation processes occurring on scales of order $\delta$ or larger. According to the present understanding of quantum turbulence, this is the case if the temperature is typically larger than $1\,K$, thanks to mutual friction \cite{Vinen2002}. 

For reference, we first note that
a similar truncation of the spectral domain has already been implemented in EDQNM simulations of the HVBK equations \cite{Tchoufag:POF2010} but, in this later study, the superfluid kinetic energy was also forced to leak out the spectral domain in such a way that the superfluid velocity spectrum would scale as $k^{-5/3}$ down to the smallest scales. This differ from our modelling, where no energy leaks to scales smaller than $\delta$.
In another related study\cite{Roche2fluidCascade:EPL2009}, an artificial superfluid viscosity was introduced in the HVBK model to force damping at small scales. This allowed to obtain an extended inertial cascade by direct numerical simulations, but at the expense of an artificial modeling of the small scales.

The computational domain is cubic (size $2\pi$) with periodic boundary conditions in the three directions. The spatial
resolution is $512^3$, unless otherwise specified. A random forcing 
(acting in the shell of wave-vectors $1.5 < |\mathbf{k}| < 2.5$) 
is imposed on the normal-fluid alone at the two highest simulated temperatures and on the superfluid alone at the two lowest ones \cite{Roche2fluidCascade:EPL2009}.

The parallel code has been adapted from an existing validated code \cite{Leveque:2001,Roche2fluidCascade:EPL2009}, based on a pseudo--spectral method with
$2^{\rm nd}$ order accurate Adams--Bashforth time stepping. Usual checks have been done on the simulations (solenoidal condition, balance of the various energy fluxes, robustness to the anti-aliasing procedure). We also checked that the normal-fluid is well damped at cut-off wave-vector, in other words that our truncation procedure, which is motivated by the superfluid physics but imposed to both fluids, is not biasing the normal fluid dynamics.
 Finally, the more complete HVBK coupling term Eq. \ref{eq:fnsHVBK} was also implemented to make sure that second order contributions to the mutual coupling don't alter the conclusion of this study.

Calculations were performed with density ratios $\rho_s/\rho_n$  $0.1$ ($\sim 2.1565$ K), $1$ ($\sim 1.96$ K), $10$ ($\sim 1.44$ K) and $40$ ($\sim 1.15$ K). The corresponding temperatures \cite{DonnellyBarenghi1998}
will be referred as \textit{high}, \textit{intermediate}, \textit{low} and\textit{ very-low}
. To simplify the analysis, we set $B=2$ (unless otherwise specified) omitting a two-fold temperature dependence of this parameter.


\section{Equipartition of superfluid energy at mesoscales}

\begin{figure}
\center
\includegraphics[width=1\linewidth]{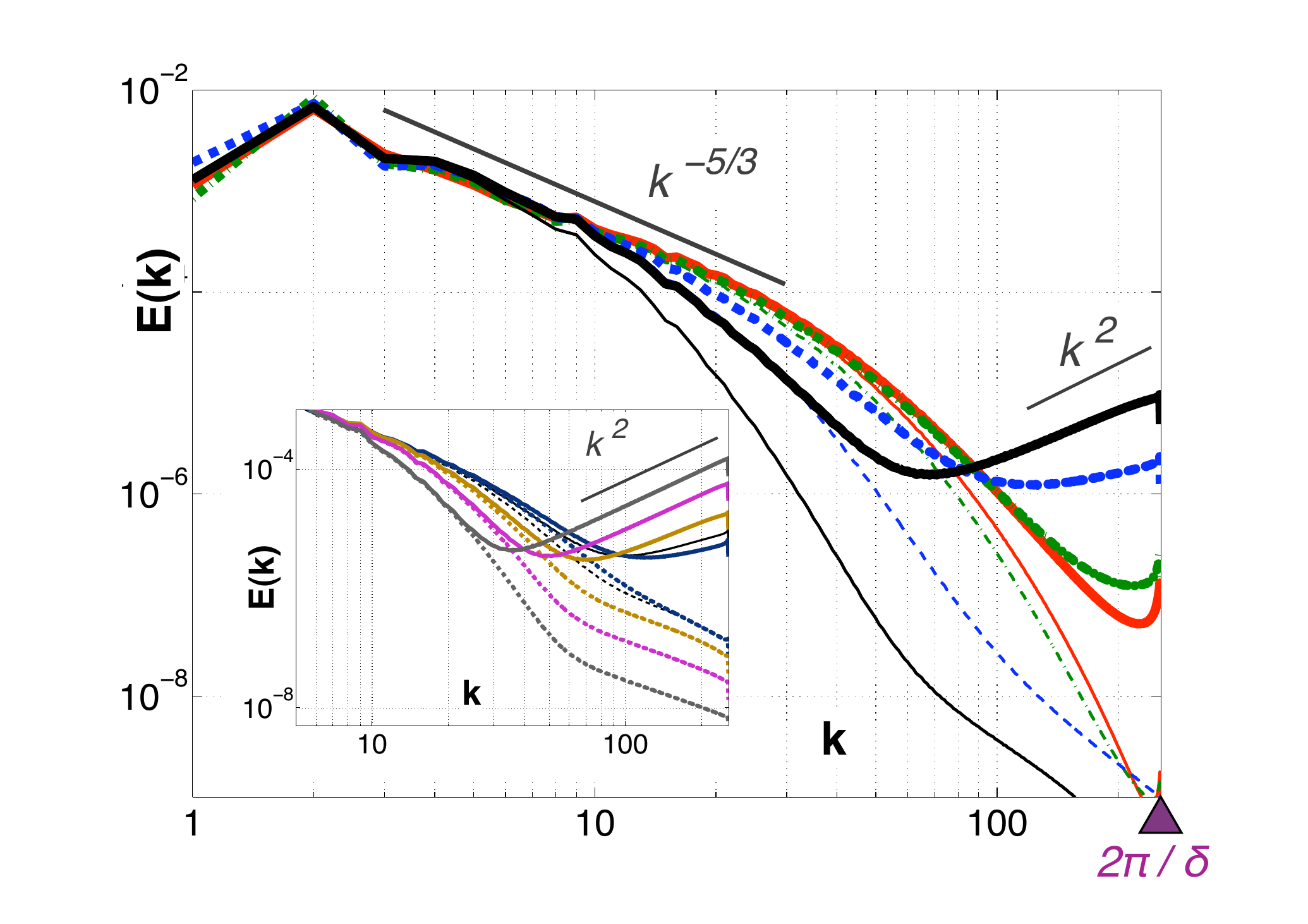}
\caption{[Coloured online] Velocity power spectra of the superfluid (thick lines) and normal-fluid (thin lines) at high (red solid line), intermediate (green dash dotted line) and low (blue dashed line) and very low (black solid line) temperatures. The wavenumber $2\pi / \delta \simeq 256$ associated with the intervortex spacing is marked by a triangle. Insert : similar spectra at low temperature only for different strength of the mutual coupling constant $B$. From top to bottom, the thick solid lines correspond to $B=0.02$ (red), $0.1$ (purple), $0.5$ (blue), $1$ (green). The thin black lines corresponds to a more complete expression of the HVBK mutual coupling (Eq. \ref{eq:fnsHVBK} with $B=2$ and $B^\prime =0.6$).
} 
\label{spectreV} 
\end{figure}

Fig. \ref{spectreV} presents the velocity spectra of the superfluid and the normal-fluid at various temperatures. At high temperature, where the normal-fluid is dominant ($\rho_s / \rho _n=1/10$), we recognize the inertial spectra, with a $-5/3$ scaling at large scales and a sharper and sharper cut-off at small scales. As the temperature decreases, the $-5/3$ scaling remains present at the largest scales in agreement with experimental \cite{Maurer1998,Salort:PoF2010} and numerical  \cite{Merahi:2006,Roche2fluidCascade:EPL2009} findings in such conditions, but a new behaviour appears at small scales. A first feature of this new behaviour is the upward inflection of the normal-fluid spectra (e.g. for $k\simeq 60$ on the thin black line of Fig. \ref{spectreV}), which contrasts with the exponential decay for ordinary fluids. A second feature is the increase of superfluid spectra versus $k$. The scaling of the superfluid spectra tends toward $k^2$ at the lowest temperature, where the superfluid is dominant. Such a scaling is typical of equipartition of energy among the hydrodynamic modes  \cite{Orszag_LesHouches1973}. The range of scales over which the system exhibits this new behaviour will be called the \textit{mesoscales}, as it sits between the large scales of inertial cascade and the microscopic scales associated with individual quantum vortices. We note that an accumulation of superfluid excitations at small scales has already been predicted \cite{RocheInterpretation:EPL2008} to interpret experimental results \cite{RocheVortexSpectrum:EPL2007}.

Physically, the emergence of this range of mesoscales at low temperature results from difficulty of dissipating superfluid kinetic energy at the bottom of the inertial cascade. We already pointed that the mutual coupling force 
 $ ({B \rho_n}/{2 \rho} ) \vert \bom \vert \vns$ is proportional to the normal-fluid fraction, which tends to zero at low temperature. As a result, the superfluid kinetic energy piles up at small scales, leading -in return- to an increase of the $\vert \bom \vert$ and $\vns$ factors in the mutual coupling. The resulting stationary equilibrium exhibits a mesoscale bath of superfluid excitations.

To verify that the trend to equipartition only results from weakening of the coupling term, and not indirectly from other effects, we artificially reduced by steps the mutual coupling coefficient $B$ keeping all other parameters constant\footnote{in particular, we no longer impose that the numerical resolution cut-off matches the estimated quantum cut-off, which varies from $2\pi/\delta=781$ to 1507.}. This procedure also allows to check that the $k^2$ scaling is truly  the asymptotic limit of the mesoscale superfluid spectrum. 
The resulting spectra, obtained for $\rho_s / \rho_n =10$, are presented in the insert of figure \ref{spectreV}. They confirm the trend to the equipartition $k^2$ scaling in the limit of low coupling. This insert also presents simulations carried out with the more sophisticated coupling term, matching the complete HVBK expression with zero vortex tension. The emergence of mesoscales is found to be robust to such a change of the coupling term.

For reference, we first note that simultaneous observation of a $k^{-5/3}$ inertial cascade and a $k^{2}$ equipartition at small scales have already been reported in \textit{truncated Euler} simulations (e.g. \cite{Cichowlas2005,Bos:PoF2006}). In contrast with our \textit{truncated HVBK} model,  a \textit{truncated Euler} system can never reach a stationary state due to the absence of dissipation. In recent work\cite{Krstulovic:PRL2011}, a truncated version of the Gross-Pitaevskii equation was implemented to describe turbulent Bose Einstein condensates, and a transient $k^{2}$ scaling was also evidenced at small scale. A spectrum with both a $k^{-5/3}$ and $k^{2}$ scaling was also predicted in  \cite{Lvov:2007} to account for superfluid turbulence at zero-temperature \cite{CommentaireModelLvov}. 
Finally, a stationary $k^2$ spectrum is observed in the near-dissipation range when replacing the standard viscous dissipation process by an higher-order hyperviscous dissipation in the hydrodynamical equations \cite{Frisch:2008}.

\section{The superfluid vortex bath}

\begin{figure} 
\center
\includegraphics[width=1\linewidth]{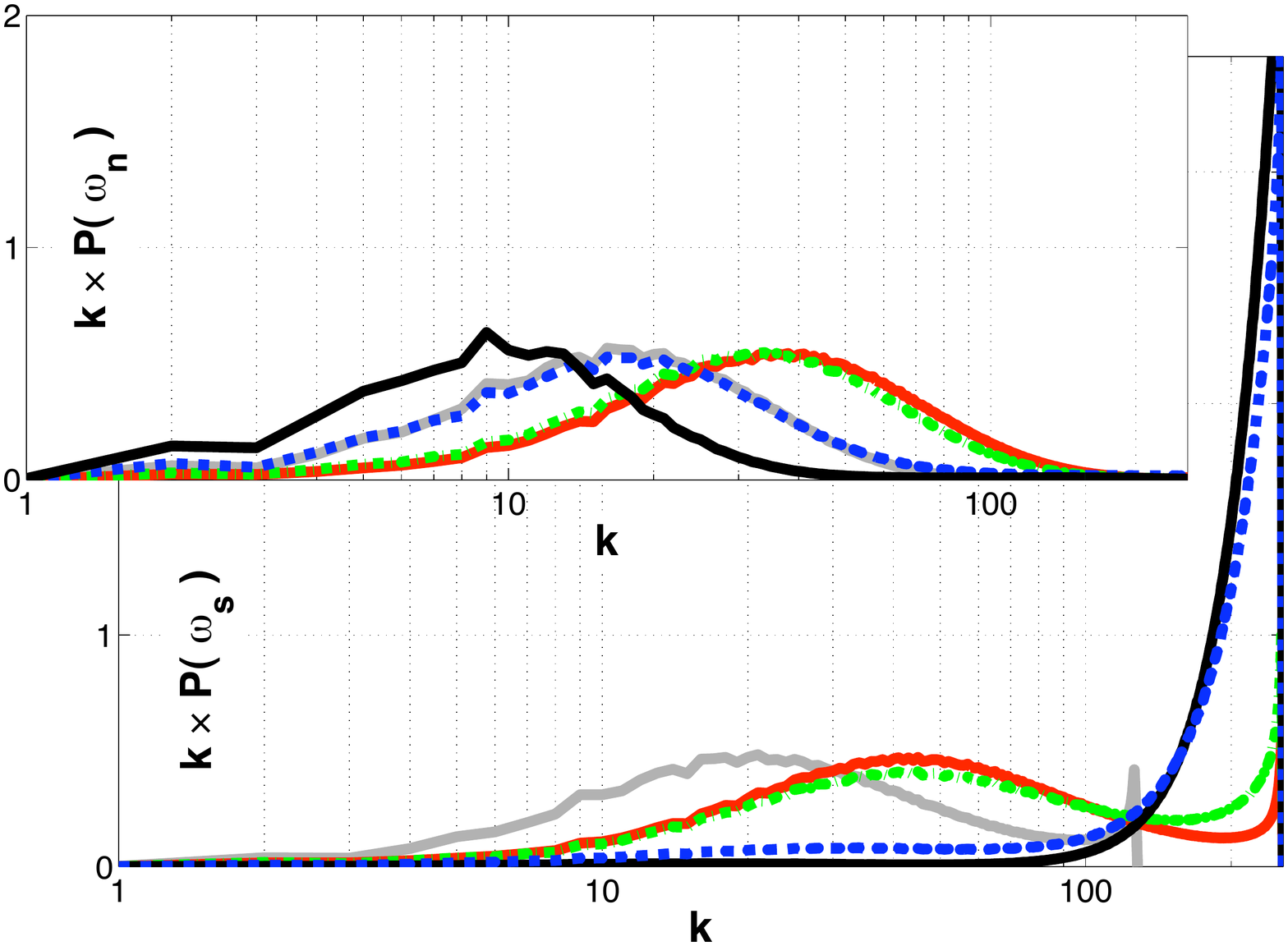} 
\caption{[Coloured online] Lower plot : Power spectral density  $P(\omega _s)$  of the superfluid vorticity.  Upper plot : corresponding quantity in the normal-fluid. The $k$ compensation on the vertical axis, associated with the log-linear representation, allows to interpret the area below the curves as the total square vorticity (enstrophy). High (red and grey lines), intermediate (green dash-dotted line), low (blue dashed line) and very low (black line) temperatures. The grey data have been obtained at lower resolution ($256^3$ instead of $512^3$) and with a reduced external forcing of the fluid such that the new cut-off wavevector $k=128$ still matches the estimated quantum cut-off scale.
} 
\label{semilogW} 
\end{figure}

The vorticity spectra are plotted in Fig. \ref{semilogW}.
This figure illustrates that most of the superfluid vorticity is concentrated at large scales at high temperature, and at small scales at low temperature\footnote{In contrast, the superfluid kinetic energy remains mostly localized in the inertial range.}. As a consequence regarding superfluid turbulence experiments, second sound measurements of the vortex line density should carry a different piece of information about the flow in each of these temperature limits.

\subsection{Inhomogeneities of the mesoscales bath}
\begin{figure*} 
\begin{center}
\center
\includegraphics[width=1\linewidth]{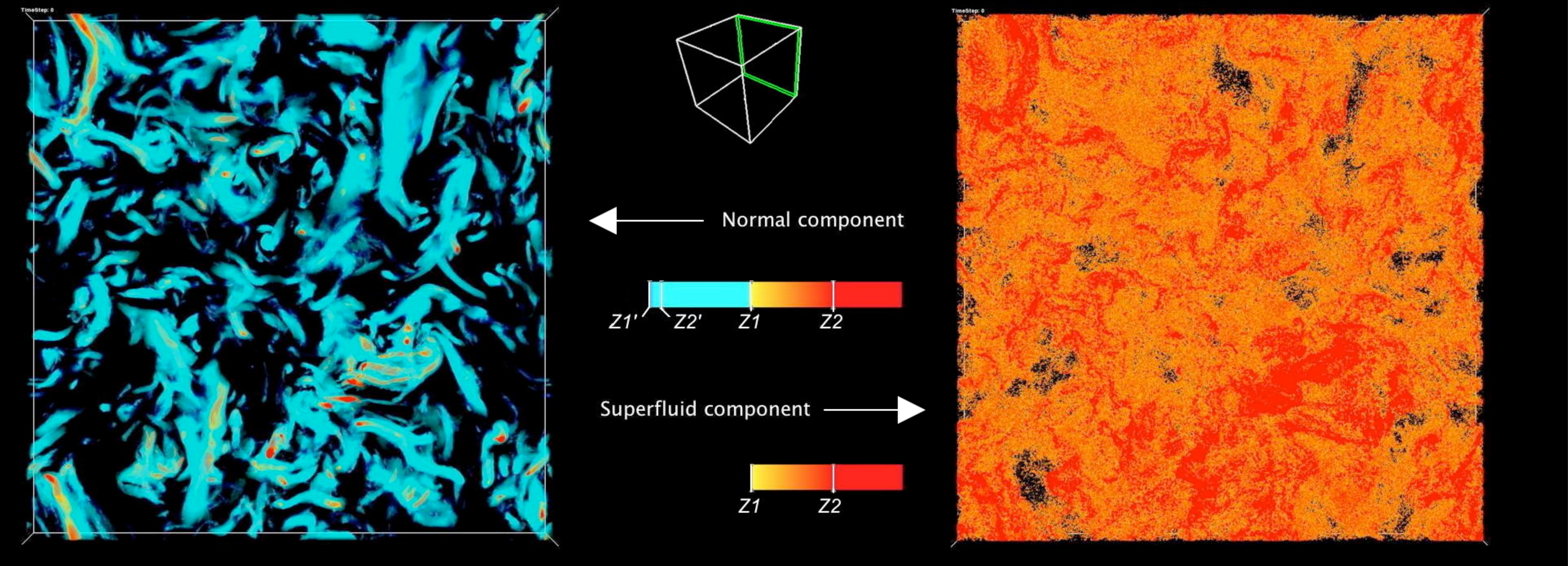} 
\caption{[Coloured online] Square vorticity field  
of normal-fluid (left) and superfluid (right) at low temperature ($\rho_n=\rho_s /10$). The in-between sketch  shows the simulation-box (white) and the slice of fluid being displayed (green). The coloring evidences the regions with the highest vorticity.
The color threshold $Z2$ (resp. $Z1$) is such that 75\% (85\%) of the superfluid total square vorticity is concentrated in the $Z_s > Z2$ (resp. $Z_s > Z1$) region. The threshold $Z2^\prime$ (resp. $Z1^\prime$) is such that 75\% (85\%) of the normal total square vorticity is concentrated in the $Z_n > Z2^\prime$ (resp. $Z_n > Z1^\prime$) region. (Generated with vapor [http://www.vapor.ucar.edu/])} 
\label{fig:Visu} 
\end{center} 
\end{figure*}

Fig. \ref{fig:Visu} presents a thin slice of the simulation box in the low temperature case ($\rho_n=\rho_s /10$). Using the same color code (see legend), it displays the normal-fluid (left image) and superfluid (right image) square vorticity fields. As expected from the spectra, a significant amount of vorticity has accumulated in the superfluid, compared to the normal-fluid. A less obvious result is the large-scale organization of the superfluid vorticity, which remains spatially correlated with the normal-fluid one. The mesoscale bath of superfluid excitations does not uniformly fill the flow  independently of the large scale dynamics of the flow.

\section{Intervortex spacing versus Reynolds number}

\begin{figure}
\center
\includegraphics[width=1\linewidth]{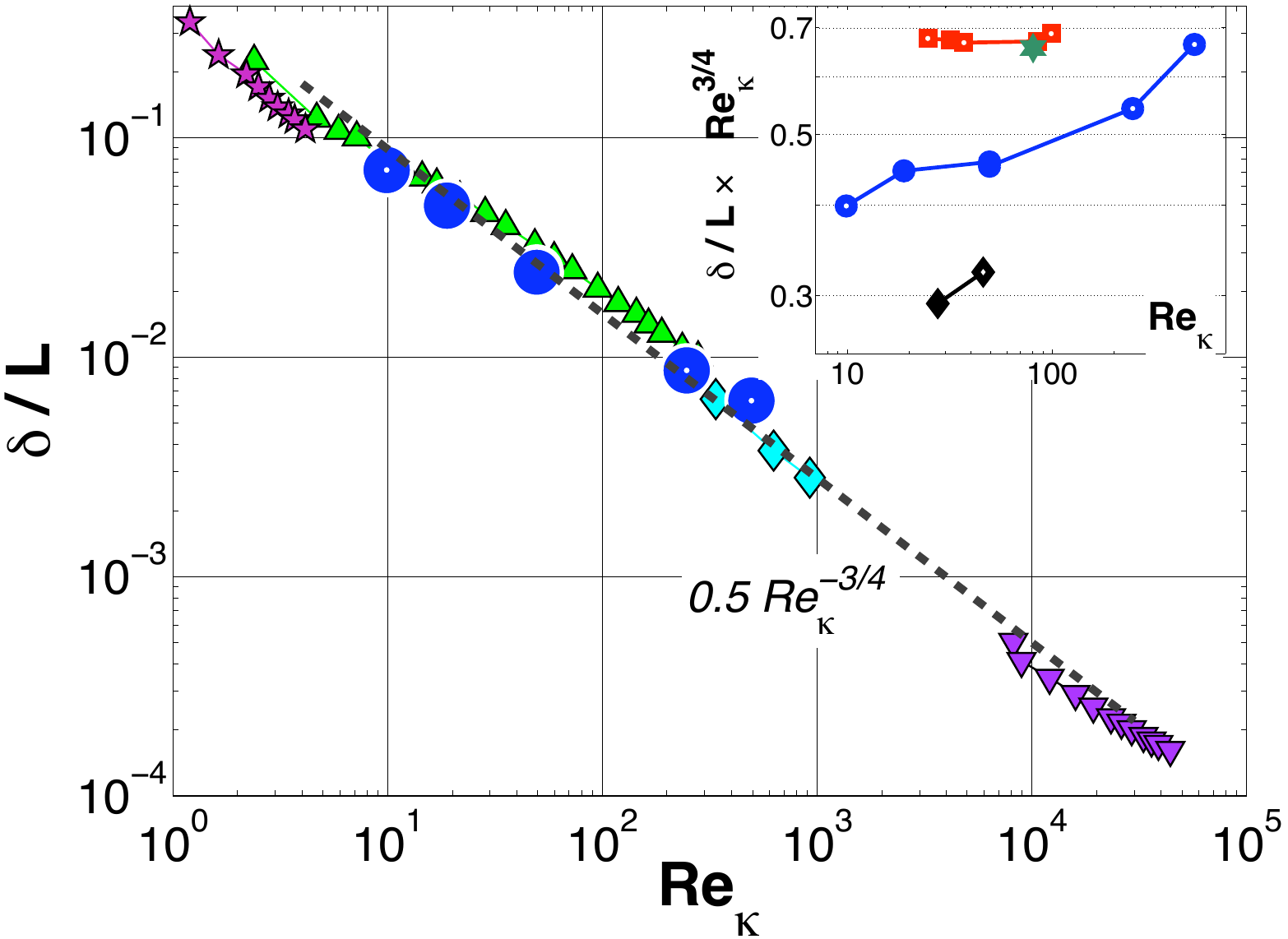} 
\caption{[Coloured online] Intervortex spacing $\delta$ (or its numerical ansatz) normalised by the integral scale $L$ versus $Re_{\kappa}$.   Present simulations at low temperature ($T\simeq 1.44\,K$) : blue discs. Experimental results within $~1.5-1.6$\,K : purple stars \cite{Ijsselstein1979}, chartreuse pointing-up triangles \cite{Holmes1992}, turquoise diamonds \cite{Walstrom1988b}, lilac pointing-down triangles : unpublished measurements obtained with the apparatus and probe described in \cite{RocheVortexSpectrum:EPL2007}. Dashed line : fit.
Insert : Compensated intervortex spacing for the present simulations at high (red squares), intermediate (green star), low (blue discs) and very low (black diamonds) temperatures.
}
\label{VersusRe} 
\end{figure}

It is tempting to relate the simulated superfluid vorticity 
and the quantum vortex line density $\mathcal{L}$ measured in experiments. To do so, we compiled measurements of $\mathcal{L}$ obtained in various channel and pipe flows within a narrow temperature range ($1.5<T<1.6\,K$ corresponding to $7.6 > \rho_s / \rho_n>4.9$) (\cite{Ijsselstein1979,Holmes1992,Walstrom1988b,RocheVortexSpectrum:EPL2007}).
 We estimate the mean inter-vortex spacing in experiments  as $\delta = {\mathcal{L}}^{-1/2}$ and in simulations from the vorticity using Eq. \ref{eq:truncation}. Then, this spacing $\delta$ is made dimensionless using the integral length scale $L$ of the flows\footnote{when $L$ is not available experimentally, it is estimated as half of the flow channel/pipe width}. In Fig. \ref{VersusRe}, $\delta / L$ is plotted against the Reynolds numbers :
\begin{equation}
Re_{\kappa}=\frac{L V_{rms}} { \kappa} 
\end{equation}

where 
$V_{rms}$ is the root-mean-square velocity of the flow. When $V_{rms}$ was not available experimentally (\cite{Ijsselstein1979,Holmes1992,Walstrom1988b}), it was estimated as $5\%$ of the mean velocity in the channel/pipe. Fig. \ref{VersusRe} evidences a excellent agreement between the simulations and the experiments, both in magnitude and scaling versus the Reynolds numbers. At first, the quality of this agreement may be surprising, given the arbitrariness of the numerical truncation  $k \le 2\pi / \delta$. In fact, the value of $\vert \bom \vert$, and therefore $\delta$, adjusts itself to dissipate the cascade of superfluid energy by mutual friction. For a given power supply at large scales, $\vert \bom \vert$ remains independent -at first order- of the precise truncation criteria. After having checked this property, we used it to extend the range of $\delta / L$ explored numerically. This result justifies \textit{a-posteriori} the use of Eq. \ref{eq:truncation} to truncate our HVBK model. A fit at 1.6\,K is also plotted  :
\begin{equation}
\delta / L \simeq 0.5 Re_{\kappa}^{-3/4}
\end{equation}

It is interesting to note that the power law scaling $Re_{\kappa}^{-3/4}$ is similar to the one found in classical turbulence for the Kolmogorov dissipative scale, made dimensionless with the integral scale of the flow. To the best of our knowledge, this result has never been reported. The insert of the figure gathers simulations results of $\delta / L$ compensated by $Re_{\kappa}^{-3/4}$. It suggest that -at a given $Re_{\kappa}$-  the intervortex spacing increases with the temperature till a saturation which roughly corresponds to the disappearance of the range of mesoscales (see spectra). As expected from the analysis of Fig. \ref{semilogW}, the intervortex spacing carries two types of information : one about the inertial cascade and the other one about the mesoscales. We recall that the mutual coefficient $B$ was taken constant in these simulations : its (weak) temperature dependence should be taken into account to predict more precisely temperature dependence of $\delta / L$ at given $Re_{\kappa}$.


\section{The vortex density spectrum}

One of the unexpected experimental results in superfluid turbulence is related to the spectrum of the vortex line density. Local measurements performed near 1.6\,K with a miniature second sound probe evidenced a decreasing spectrum at large scales\cite{RocheVortexSpectrum:EPL2007} with a scaling behavior close to a $k^{-5/3}$ (see also the possibly related study \cite{Bradley:PRL2008}). 
In classical turbulence, the one dimensional spectrum of the absolute value of the vorticity is rather flat or slightly decreasing\cite{RocheInterpretation:EPL2008}. This results suggests that both superfluid and classical turbulence can be distinguished from hydrodynamics measurement performed at large scales. This contrasts with existing experimental results which have always been interpreted assuming that both types of turbulence were identical at large scales.

The interpretation proposed in \cite{RocheVortexSpectrum:EPL2007} for the vortex density $-5/3$ spectrum at 1.6\,K predicts the existence of a unpolarised bath of superfluid excitations at low scales and predicts that it contributes to most of the vortex line density signal. It then predicts that the unpolarised bath is inhomogeneous and -to some extent- advected by the large scale flow. This leads to the characteristic $k^{-5/3}$ power spectrum reminiscent of passive-scalar turbulence\cite{Warhaft:2000}. Importantly, the present simulations confirms several of these qualitative predictions, in particular the existence of an ``unpolarised bath'' which corresponds to the mesoscale excitations.

\begin{figure} 
\includegraphics[width=1\linewidth]{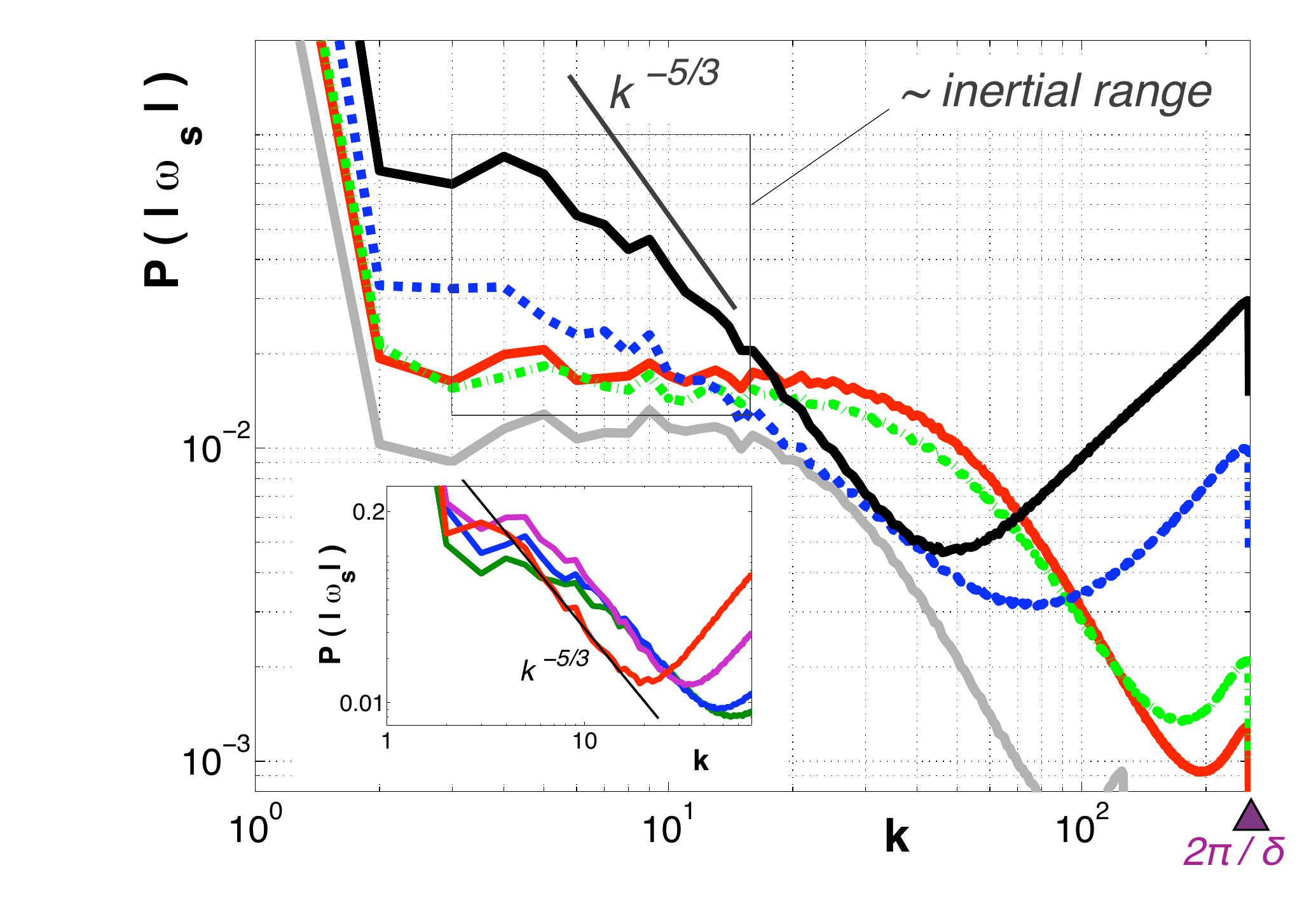} 
\caption{[Coloured online] Power spectral density of the modulus of the superfluid vorticity at high (red and grey lines), intermediate (green dot-dashed line) and low (blue dashed line) and very low (black line) temperatures. The wavenumber $2\pi / \delta \simeq 256$ associated with the intervortex spacing is marked by a triangle. Insert : similar spectra at low temperature only for different strength of the mutual coupling constant $B$ (same colour code as in Fig. \ref{spectreV}).} 
\label{spectreVorticite} 
\end{figure}

Fig. \ref{spectreVorticite} reports the simulations spectra of $\vert \bom \vert$ at different temperatures using the same color code as Fig. \ref{spectreV}. At large scales, we find that the spectrum is rather flat at high temperature, as expected in classical turbulence, but becomes more and more decreasing as the temperature gets lower, without exactly reaching the $-5/3$ power law in the present conditions (Reynolds numbers in simulations are typically 2 decades smaller than in experiments). A $k^{-5/3}$ scaling is better evidenced when decreasing the mutual force constant $B$ (see insert). This result is therefore qualitatively consistent with the experimental results. 
Interestingly, it seems that this $k^{-5/3}$ scaling develops in an intermediate range of scales between the inertial scales and the so-called meso-scales, therefore suggesting that interesting dynamics (possibly related to passive-scalar dynamics) may happen in this range of scales. This requires further investigations, in particular through simulations at larger Reynolds numbers allowing for a better separation of different ranges of scales.

\section{Perpective}

It would be interesting to confirm experimentally the existence of a range of mesoscales with equipartition of energy. This test is presently difficult with the state-of-the-art instrumentation, as the smallest velocity probes ever operated in superfluid ($\sim 500 \,\mu m$ in \cite{Salort:PoF2010}) are too large. Nevertheless, observation of these mesoscales is certainly within reach of a dedicated micro-machined probe operated in a large enough flow.

\acknowledgments 

We thank C. Barenghi, B. Castaing, N. Rousseau and P. Diribarne for their inputs and G. Gil for support in porting code to the Jade supercomputer at CINES (Montpellier, France). This work was made possible by the ANR SHREK grant and by using local computing facilities from PSMN at ENS Lyon and HPC resources from GENCI-CINES (Grant 2010-026380 and 2011-026380).



\end{document}